%% file: main.tex
\pdfoutput=1

\documentclass[sigconf,dvipsnames]{acmart}

\AtBeginDocument{%
  \providecommand\BibTeX{{%
    \normalfont B\kern-0.5em{\scshape i\kern-0.25em b}\kern-0.8em\TeX}}}

\usepackage{hyperref}
\hypersetup{bookmarksnumbered,unicode}
\usepackage{booktabs}
\usepackage{caption}
\usepackage{subcaption}
\usepackage{graphicx}
\usepackage{color,soul}
\usepackage{multirow}

\setcopyright{none}

\author{Chlo{\"e} Brown}
\authornote{Ordered alphabetically, equal contribution.}
\affiliation{%
  \institution{University of Cambridge}
 }

\author{Jagmohan Chauhan}
\authornotemark[1]
\affiliation{%
  \institution{University of Cambridge}
}

\author{Andreas Grammenos}
\authornotemark[1]
\additionalaffiliation{
    \institution{Alan Turing Institute}
}
\affiliation{%
    \institution{University of Cambridge}
}

\author{Jing Han}
\authornotemark[1]
\affiliation{
    \institution{University of Cambridge}
}

\author{Apinan Hasthanasombat}
\authornotemark[1]
\affiliation{%
    \institution{University of Cambridge}
}

\author{Dimitris Spathis}
\authornotemark[1]
\affiliation{%
    \institution{University of Cambridge}
}

\author{Tong Xia}
\authornotemark[1]
\affiliation{
    \institution{University of Cambridge}
}

\author{Pietro Cicuta}
\affiliation{%
  \institution{University of Cambridge}
}

\author{Cecilia Mascolo}
\affiliation{
  \institution{University of Cambridge}
}

\begin{document}

\title[Exploring Automatic Diagnosis of COVID-19 from Crowdsourced Respiratory Sound Data]{Exploring Automatic Diagnosis of COVID-19 \\
from Crowdsourced Respiratory Sound Data}

\renewcommand{\shortauthors}{Brown C. et al.}
\begin{abstract}
 Audio signals generated by the human body (e.g., sighs, breathing, heart, digestion, vibration sounds) have routinely been used by clinicians as indicators to diagnose disease or assess disease progression. Until recently, such signals were usually collected through manual auscultation at scheduled visits. Research has now started to use digital technology to gather bodily sounds (e.g., from digital stethoscopes) for cardiovascular or respiratory examination, which could then be used for automatic analysis. 
 Some initial work shows promise in detecting diagnostic signals of COVID-19 from voice and coughs. In this paper we describe our data analysis over a large-scale crowdsourced dataset of respiratory sounds collected to aid diagnosis of COVID-19. We use coughs and breathing to understand how discernible COVID-19 sounds are from those in asthma or healthy controls. Our results show that even a simple binary machine learning classifier is able to classify correctly healthy and COVID-19 sounds. %
 We also show how we distinguish a user who tested positive for COVID-19 and has a cough from a healthy user with a cough, %
 and users who tested positive for COVID-19 and have a cough from users with asthma and a cough. Our models achieve an AUC of above 80\% across all tasks. %
These results are preliminary and only scratch the surface of the potential of this type of data and audio-based machine learning. This work opens the door to further investigation of how automatically analysed respiratory patterns could be used as pre-screening signals to aid COVID-19 diagnosis.

\end{abstract}


\maketitle

\input{introduction}

\input{related}
\input{datacollection}
\input{method}

\input{evaluation}

\input{discussion}

\begin{acks} 
This work was supported by ERC Project 833296 (EAR) and the UK Cystic Fibrosis Trust. %
We thank A. Barnea, M.E. Bryan, and A. Floto 
for the useful discussions, and everyone who volunteered their data. %
\end{acks}

\bibliographystyle{ACM-Reference-Format}
\bibliography{refs.bib}

\cleardoublepage
\appendix
\input{appendix}

\end{document}

%% file: introduction.tex
\section{Introduction}

Audio signals generated by the human body (e.g., sighs, breathing, heart, digestion, vibration sounds) have often been used by clinicians and clinical researchers in diagnosis and monitoring of disease. However, until recently, such signals were usually collected through manual auscultation at scheduled visits. Research has now started to use digital technology to gather bodily sounds (e.g., digital stethoscopes) and run automatic analysis on the data~\cite{PBR17}, for example for wheeze detection in asthma~\cite{LLT17,OB16}. Researchers have also been piloting the use of human voice to assist early diagnosis of a variety of illnesses: Parkinson’s disease correlates with softness of speech (resulting from lack of coordination of the vocal muscles)~\cite{SGO17,BMG17}, voice frequency with coronary artery disease (hardening of the arteries which may affect voice production)~\cite{MSO18}, and vocal tone, pitch, rhythm, rate, and volume correlate with invisible illnesses such as post-traumatic stress disorder~\cite{BIX19}, traumatic brain injury and psychiatric conditions~\cite{FBF16}. The use of human-generated audio as a biomarker for various illnesses offers enormous potential for early diagnosis, as well as for affordable solutions which could be rolled out to the masses if embedded in commodity devices. This is even more true if such solutions could monitor individuals throughout their daily lives in an unobtrusive way.

Recent work has started exploring how respiratory sounds (e.g., coughs, breathing and voice) collected by devices from patients tested positive for COVID-19 in hospital differ from sounds from healthy people.  In~\cite{Huang2020.04.07.20051060}
digital stethoscope data from lung auscultation is used as a diagnostic signal for COVID-19; in~\cite{imran2020ai4covid19} a study of detection of coughs related to COVID-19 collected with phones is presented using a cohort of 48 COVID-19 patients versus other pathological coughs on which an ensemble of models are trained. In~\cite{han2020early}
speech recordings from hospital patients with COVID-19 are analyzed to categorize automatically the health state of patients. %
Our work contains an exploration of using human respiratory sounds as diagnostic markers for COVID-19 in crowdsourced, uncontrolled data. 
Specifically, this paper describes our preliminary findings over a subset of our dataset currently being crowdsourced worldwide at  \url{www.covid-19-sounds.org}.  The dataset was collected through an app (Android and Web) that asked volunteers for samples of their voice, coughs and breathing as well as their medical history and symptoms. The app also asks if the user has tested positive for COVID-19. To date, we have collected on the order of 10,000 samples from about 7000 unique users. While other efforts exist that collect some similar data, they are often either limited in scope (e.g., collect only coughs~\cite{detectnow,coughvid}) or in scale (e.g., collect smaller samples in a specific region or hospital). This is, to our knowledge, the largest uncontrolled, crowdsourced data collection of COVID-19 related sounds worldwide. In addition, the mobile app gathers data from single individuals up to every two days, allowing for potential tracking of disease progression. This is also a unique feature of our collected dataset. Section~\ref{data} contains a more detailed description of the data. In this paper we analyze a subset of our data as described in Section~\ref{subset-data} and show some preliminary evidence that cough and breathing sounds could contain diagnostic signals to discriminate COVID-19 users from healthy ones; we further compare COVID-19 positive user coughs with healthy coughs, as well as those from users with asthma. More precisely, the contributions of this paper are:

\begin{itemize}
\item Description of COVID-19 sound collection framework through apps, and the types of sounds harvested through crowdsourcing.
\item Illustration of the large-scale dataset being gathered. To date, this is the largest being collected and among the most inclusive in terms of types of sounds. It contains sounds from about 7000 unique users (more than 200 of whom reported a recent positive test for COVID-19).%
\item  We present initial findings around the discriminatory power of coughs and breath sounds for COVID-19. We construct three binary tasks, one aiming to distinguish COVID-19 positive users from healthy users; one aiming to distinguish COVID-19 positive users who have a cough from healthy users who have a cough; and one aiming to distinguish COVID-19 positive users with a cough from users with asthma who report having a cough. The results show that the performance remains above 80\% Area Under Curve (AUC) for all tasks. Specifically, we are able to classify correctly healthy and COVID-19 sounds with an AUC of 80\% (Task 1). When trying to distinguish a user who tested positive for COVID-19 and has a cough from a healthy user with a cough (Task 2), our classifier achieves an AUC of 82\%, while if we try to distinguish users who tested positive for COVID-19 and have a cough from users with asthma and a cough (Task 3) we achieve an AUC of 80\%. 
\item We show how audio data augmentation can be used to improve the recall performance of some of our tasks with less data. We see a performance improvement of 5\% and 8\% for Task 2 and Task 3 respectively. %
\item Discussion of results and their potential, and illustration of a number of future directions for our analysis and for sound-based diagnostics in the context of COVID-19, which could open the door to COVID-19 pre-screening and progression detection.
\end{itemize}

%% file: related.tex
\section{Motivation and Related Work}
Researchers have long recognised the utility of sound as a possible indicator of behavior and health. Purpose-built external microphone recorders have been used to detect sound from the heart or the lungs using stethoscopes, for example. These often require listening and interpretation by highly skilled clinicians, and are recently and rapidly being substituted by different technologies such as a variety of imaging techniques (e.g., MRI, sonography), for which analysis and interpretation is easier. However, recent trends in automated audio interpretation and modeling has the potential to reverse this trend and offer sound as the cheap and easily distributable alternative.

More recently the microphone on commodity devices such as smartphones and wearables have been exploited for sound analysis. In~\cite{CLL12} the  audio from the microphone is used to understand the user context and this information is aggregated to make up a view of the ambience of places around a city. In Emotionsense~\cite{rachuri2010emotionsense}, the phone microphone is used as a sensor for detecting users' emotion  in-the-wild, through Gaussian mixture models. In~\cite{NGW15} authors analyze sounds emitted while the user is sleeping,  to identify sleep apnea episodes. Similar works have also used sound to detect asthma and wheezing~\cite{LLT17,OB16}.

Machine learning methods have been devised to recognize and diagnose respiratory diseases from sounds~\cite{PBR17} and more specifically coughs: \cite{bales2020machine} uses convolutional neural networks (CNNs) to detect cough within ambient audio, and diagnose three potential illnesses (bronchitis, bronchiolitis and pertussis) based on their unique audio characteristics.

Clinical work has concentrated on using voice analysis for specific diseases: for example, in Parkinson’s disease, microphone and laryngograph equipment have been used to detect the softness of speech resulting from lack of coordination over the vocal muscles~\cite{SGO17,BMG17}. Voice features have also been used to diagnose bipolar disorder~\cite{FBF16}; and to correlate tone, pitch, rhythm, rate, and volume with signs of invisible conditions like post traumatic stress disorder~\cite{BIX19}, traumatic brain injury and depression. Voice frequency has been linked to coronary artery disease (resulting from the hardening of the arteries which may affect voice production)~\cite{MSO18}. Companies such as Israeli-based Beyond Verbal and the Mayo Clinic have  indicated in  press releases that they are piloting these approaches.

Recently, with the advent of COVID-19, researchers have started to explore if respiratory sounds could be diagnostic~\cite{deshpande2020overview}. In~\cite{Huang2020.04.07.20051060}
digital stethoscope data from lung auscultation is used as a diagnostic signal for COVID-19.
In~\cite{imran2020ai4covid19} a study of detection of coughs related to COVID-19 is presented using a cohort of 48 COVID-19 tested patients versus other pathological coughs, on which an ensemble of model are trained.
In~\cite{han2020early} speech recordings from COVID-19 patients are analyzed to categorize automatically the health state of patients from four aspects, namely severity of illness, sleep quality, fatigue, and anxiety. Quatieri {\it et al.}~\cite{quatieri2020framework} showed that changes in vocal patterns could be a potential biomarker for COVID-19.  

Our work differs from these works, as we use an entirely crowdsourced dataset, for which we must trust that the ground truth is what the users state (in terms of symptoms and COVID-19 testing status); we must further overcome the challenges of data coming from different phones and microphones, possibly in very different environments. The closest to our work is~\cite{imran2020ai4covid19}, from which our approach differs  in two significant ways. Firstly, their data is collected in  a controlled setting. In comparison our data is crowdsourced, making data analysis more challenging. Secondly, they used an end-to-end deep learning model on their dataset consisting of around 100 samples; %
deep learning models often overfit on such very small datasets, so we chose a different strategy.   We use  simple machine learning models such as SVM  with  various features (handcrafted and obtained through transfer learning) and data augmentation to overcome such issues.  %
Other crowdsourced approaches of this kind are starting to emerge: in~\cite{sharma2020coswara} a web form to gather sound data is presented, which collected about 570 samples but does not report any COVID-19 detection analysis. Our app collected samples from more than 7000 unique users with more than 200 positive for COVID-19, and allows users to go back to the app after a few days to report progression and give another sample. We report our preliminary findings which suggest that sounds could be used to inform automatic COVID-19 screening.

%% file: datacollection.tex
\section{Data Collection}
\label{data}
This section describes our data collection framework and some properties of the gathered data. We further describe in detail the subset of the data used for the analysis in this paper. We note that the data collection and study have been approved by the Ethics Committee of the Department of Computer Science and Technology at the University of Cambridge.

\begin{figure} [t]
\centering
\begin{subfigure}{.23\textwidth}
  \centering
  \includegraphics[height=6cm, width = 3.25cm]{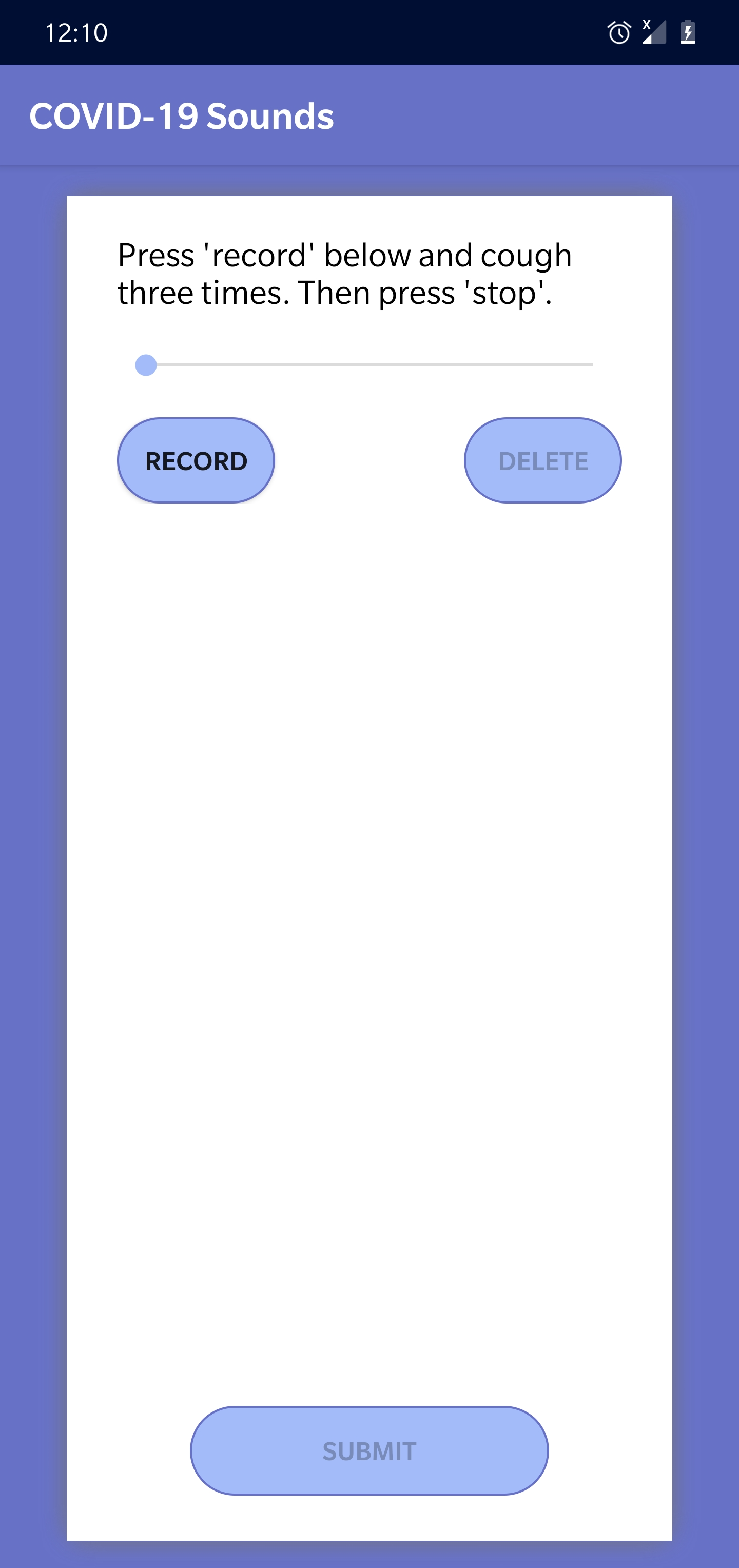}
  \caption{Cough Screen}
\end{subfigure}
\begin{subfigure}{.23\textwidth}
  \centering
  \includegraphics[height=6cm,width=3.25cm]{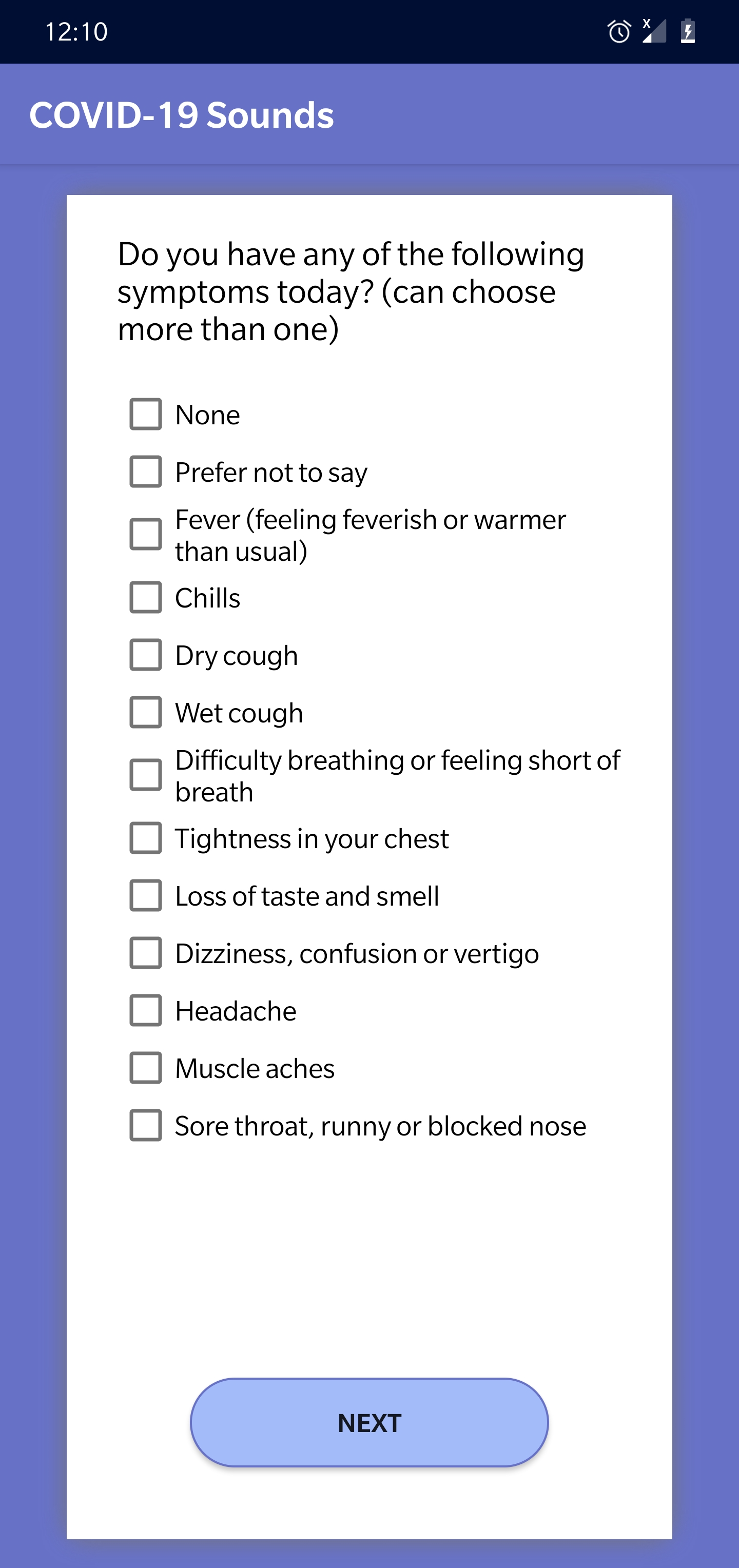}
  \caption{Symptom Screen}
\end{subfigure}

\caption{App Screens.}%
\label{fig:screens}
\end{figure}

\subsection{Our data collection apps}
Our crowdsourced data gathering framework is comprised of a web-based app and an Android app~\footnote{an iOS app is also now available at \url{www.covid-19-sounds.org}} . The features of these apps are mostly similar:
the user is asked to input their age and gender as well as a brief medical history and whether they are in hospital. Users then input their symptoms (if any) and record respiratory sounds: they are asked to cough three times, to breathe deeply through their mouth three to five times and to read a short sentence appearing on the screen three times. 
Finally, users are asked if they have been tested for COVID-19, and a location sample is gathered with permission. Figure~\ref{fig:screens} illustrates some sound- and symptom-collection screens of the Android app.
In addition, the Android (and iOS) app prompts users to input further sounds and symptoms every two days, providing a unique opportunity to study the progression of user health based on sounds. 
The data flows encrypted to our servers where it is stored securely; data is transmitted from the phones when the user is connected to WiFi and stored locally until then. If a successful transmission happens the data is removed from the device.
We do not collect user email addresses or explicit personal identifiers. 
The apps display a unique ID at the end of the survey to enable users to contact us and ask for their data deletion.  
The app does not provide medical advice to users. %
To foster reproducibility, we will release the code of our apps as open source linked from our webpage  \url{www.covid-19-sounds.org}. Given the data is sensitive (i.e., containing voice) we are setting up sharing agreements for the data.

\subsection{Crowdsourced dataset}
Helped by a large media campaign orchestrated by the University, we were able to crowdsource data from a large number of users. In particular, as of 22 May 2020, our dataset is composed of 4352 unique users collected from the web app and 2261 unique users collected from the Android app, comprising 4352 and 5634 samples respectively. Of these, 235 declared having tested positive for COVID-19: 64 in the web form and 171 in the Android app. Of the Android users, 691 users contributed more than one sample, i.e., they returned to the app after two days and reported their symptoms and sounds again.

\begin{figure} [h]
\centering
    \begin{subfigure}{.24\textwidth}
      \hspace*{-0.7cm}  
        \centering
        \includegraphics[width=\linewidth, height=5cm, keepaspectratio]{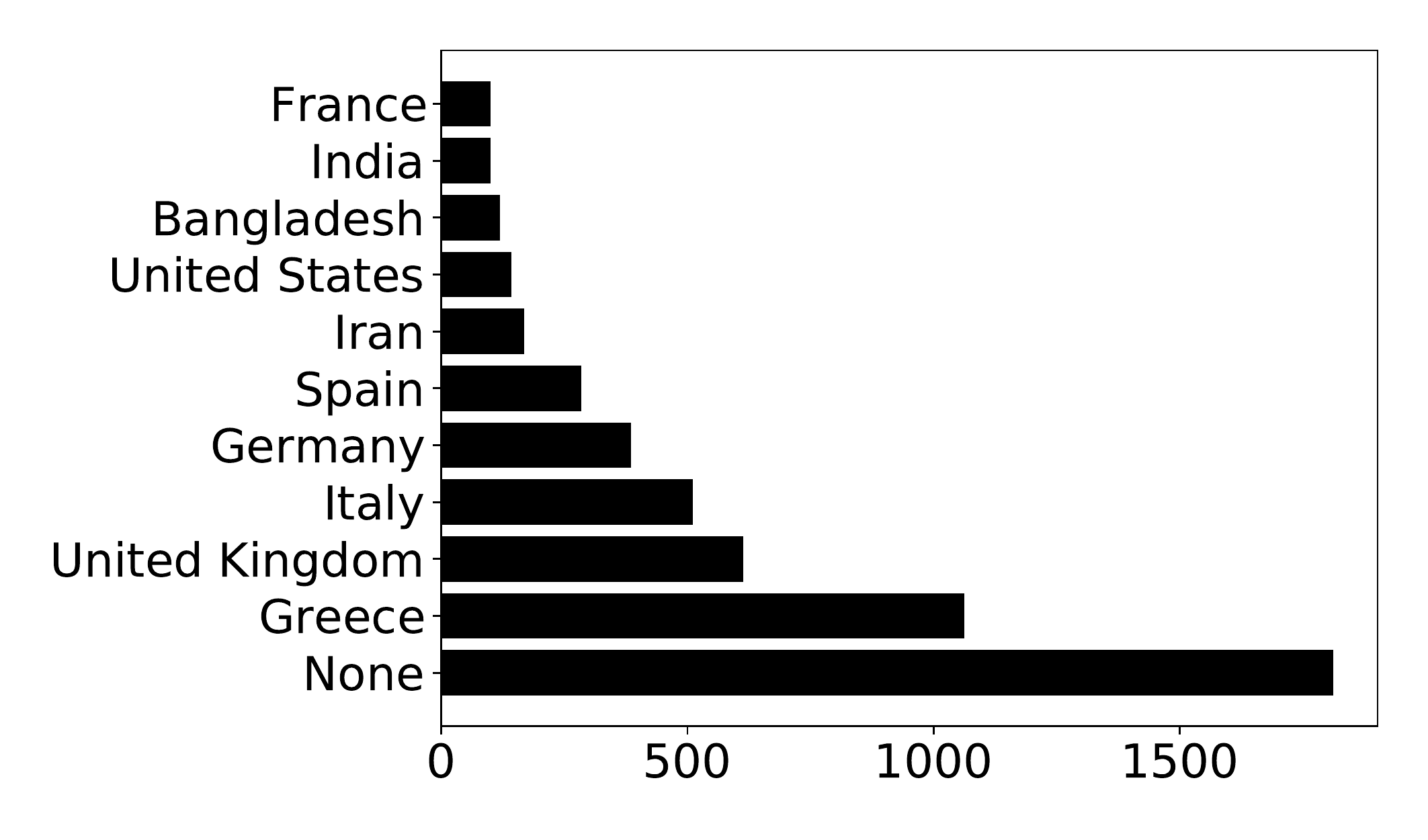}
        \label{fig:countriesDist}
        \caption{Per country}
    \end{subfigure}
    \begin{subfigure}{.23\textwidth}
      \centering
        \includegraphics[width=\textwidth, height=5cm, keepaspectratio]{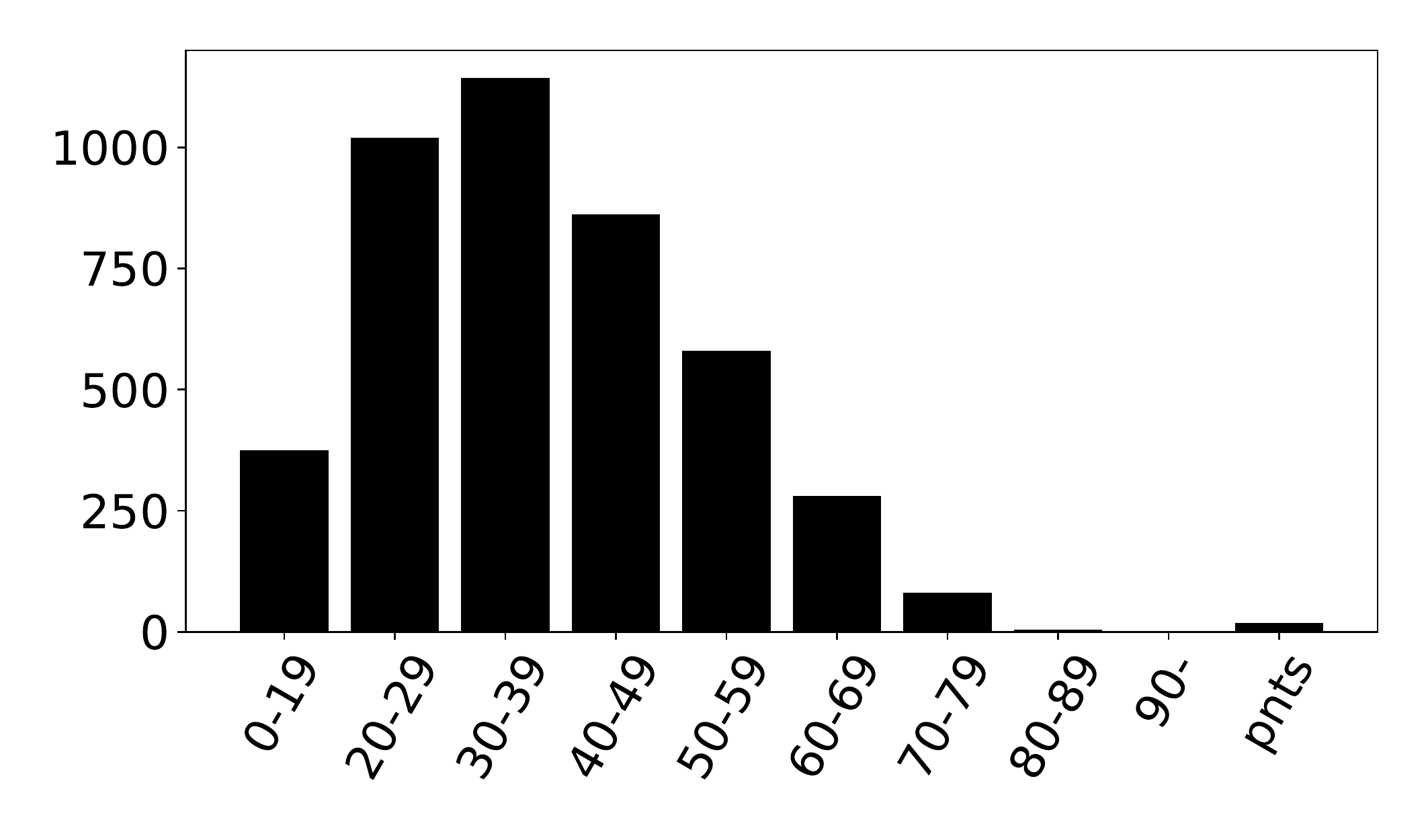}
        \caption{Age}
        \label{fig:ageDistribution}
    \end{subfigure}

\caption{User Distribution: (a) Top 10 countries, and (b) Age. pnts="Prefer not to Say", None=Country not available.
} %
\label{fig:stats}

\end{figure}

\begin{figure} [h]
  \begin{subfigure}{.49\textwidth}
    \centering
    \includegraphics[trim={0 0cm 0 0}, width=\textwidth, keepaspectratio]{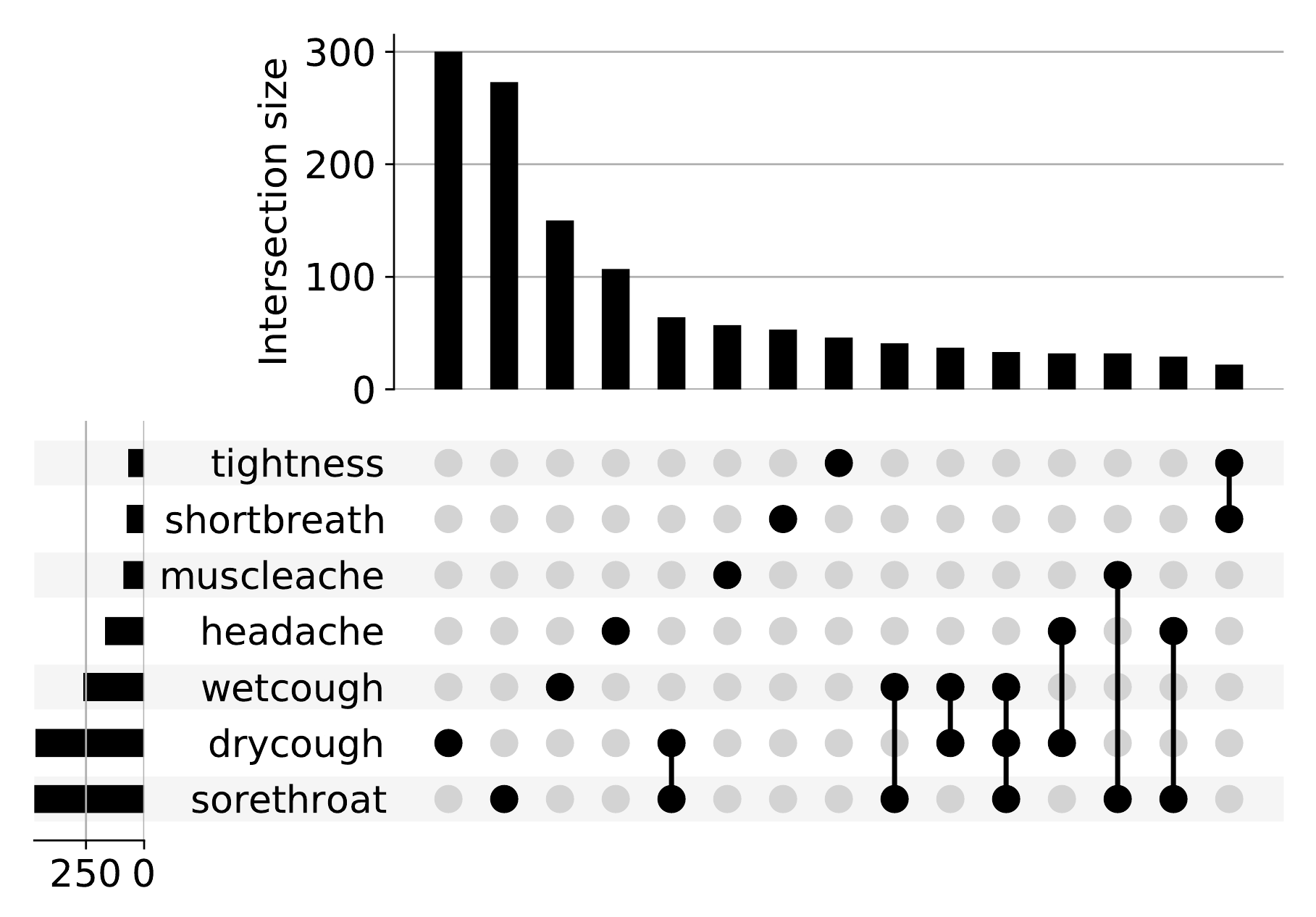}
    \label{fig:androidSymptomsDist}
    \caption{Distribution of symptoms amongst \textit{all} users.}
  \end{subfigure}
  \begin{subfigure}{.48\textwidth}
    \centering
    \includegraphics[trim={0 0cm 0 0}, width=\textwidth, keepaspectratio]{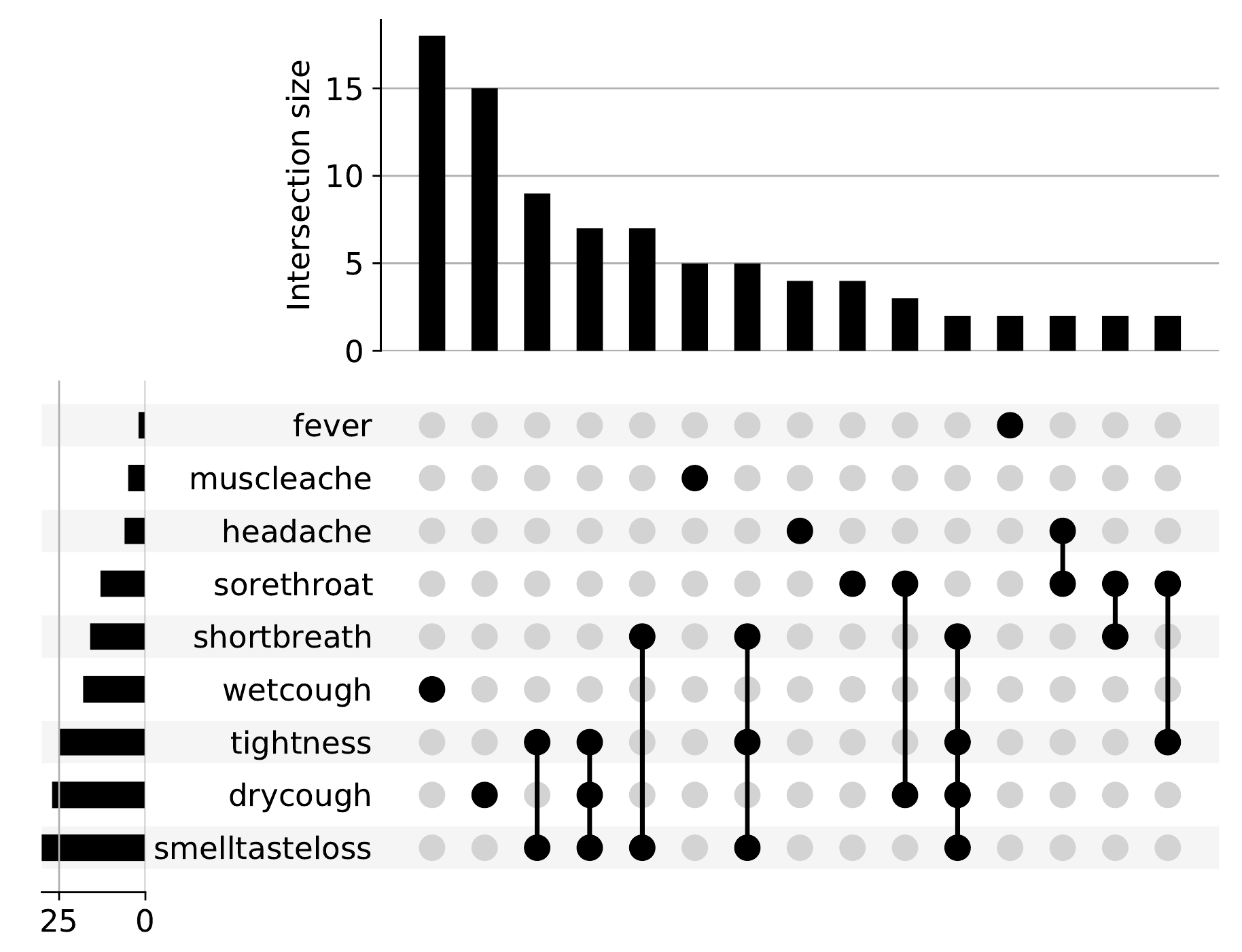}
    \label{fig:androidSymptomsDistCovidOnly}
    \caption{Distribution of symptoms in COVID-19 positive tested users.}
  \end{subfigure}

    \caption{The most frequent 15 symptom combinations obtained from Android.}
    \label{symptoms}

\end{figure}  

Data distribution statistics are described below.
Figure~\ref{fig:stats}~(a) illustrates the country (recorded from location sample) distribution. We note that many users opted not to record their location. The gender breakdown is 4525 Male, 2056 Female, 26 Prefer not to say, and six Others. Of all completed surveys, 6088 reported no symptoms and 3898 ticked at least one. Figure~\ref{fig:stats}~(b) shows the age distribution, which is skewed towards middle age.

Figure~\ref{symptoms}~(a) shows the most frequent symptom distributions for all the Android users; we do not know which users have or have had COVID-19 recently, but we know that only a small fraction of these have tested positive (see statistics above). In this group, the most common single symptom reported is a dry cough, while the most common combination of symptoms is a cough and sore throat. Figure~\ref{symptoms}~(b) shows the most frequent symptoms of users declaring a positive COVID test. Interestingly, the most common single symptoms are wet and dry cough, and the most common combination is lack of sense of smell and chest tightness. This is aligned with the COVID-19 symptom tracker data~\cite{zoe20}. The fact that cough is one of the most reported symptoms for COVID-19, but is also a general symptom of so many other diseases, provides further motivation for using sounds as a general predictor.

\subsection{Dataset used for this analysis}
\label{subset-data}
Guided primarily by the imbalance of COVID-19 tested users in the dataset, for this analysis we have focused on a curated set of the collected data (until 22 May, 2020). We also restricted our work to use only coughs and breathing (and not the voice samples). A sample is an instance of one audio recording. We report here the number of samples used in our analysis after filtering (silent and noisy samples).  In particular, we have extracted and manually checked all samples of users who said they had tested positive for COVID-19 (in the last 14 days or before that) resulting in 141 cough and breathing samples. 54 of these samples were from users who reported dry or wet cough.  %

As a control group, our analysis uses three sets of users. The first set consists of users from countries where the virus was not prevalent at the time of data collection (up to around 2000 cases): we treat these as  {\em non-COVID} users. We selected Albania, Bulgaria, Cyprus, Greece, Jordan, Lebanon, Sri Lanka, Tunisia, and Vietnam. Specifically, we define \textit{non-COVID} users as those with a clean medical history, who had never smoked, had not tested positive for COVID-19, and did not report any symptoms.  These users contributed 298 samples.  The second set \textit{non-COVID with cough} consists of users who meet the same criteria as the \textit{non-COVID}  users, but declared a cough as symptom;%
 these provided 32 samples. Finally, \textit{asthma with cough} are the users who have asthma, had not tested positive for COVID-19, and had a cough; these gave us 20 samples. 

We intend to release all our data openly; however, due to the sensitive nature (e.g. voice) our institution has advised us to release it with one-to-one legal agreements with other entities for research purposes. Our web page will include information about how to access the data.

%% file: method.tex
\section{Methods}
\label{method}

\begin{figure*} [h]
    \centering
    \includegraphics[trim={0 0.0cm 0 0}, scale=0.70]{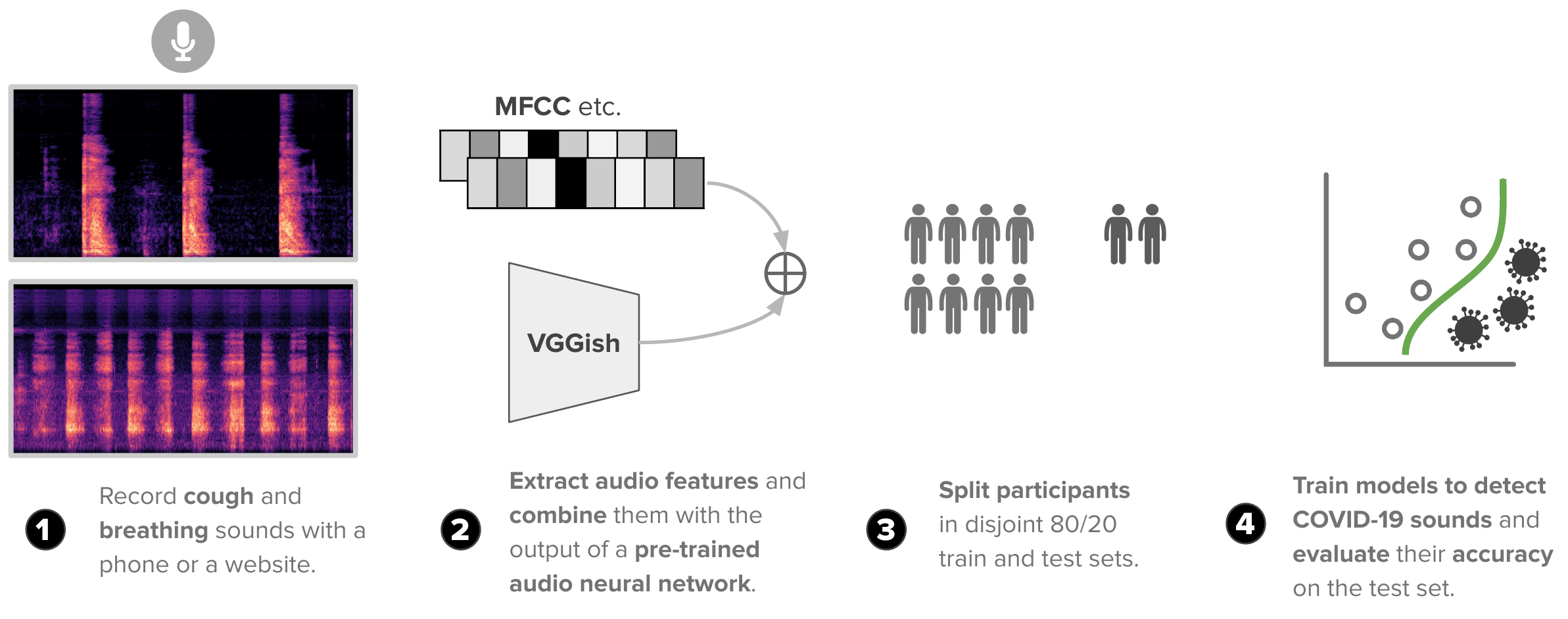}
    \caption{Description of our machine learning pipeline, describing sounds input (coughs and breathing), the extracted feature vector, and our training and testing split of the users that are used to train classification models.} 
    \label{fig:system}

\end{figure*}
Standard data processing and modeling practises from the audio and sound processing literature targeting medical applications were followed~\cite{pramono2016cough}. Based on the moderate size of the dataset selected, and the importance of explicability given the public health implications of our work, feature-based machine learning and shallow classifiers were employed. In this section, we describe the extracted features and the methodology we followed to train robust classification models, taking into account specific idiosyncrasies of our data (e.g., longitudinal mobile users and cross-validation).  We analyzed two different types of features: handcrafted features, and features obtained though transfer learning. We tested classifiers such as Logistic Regression (LR), Gradient Boosting Trees and Support Vector Machines (SVMs); results can be found in the results section.  We evaluated an SVM classifier with a Radial Basis Function (RBF) kernel. We considered different values of the following hyper-parameters: regularization parameter \textit{C} and kernel coefficient $\gamma$. Figure~\ref{fig:system} illustrates the data processing pipelines.

\subsection{Feature extraction} 

\textbf{Handcrafted Features.} The raw sound waveform recorded by the apps is resampled to 22kHz, a standard value for audio tasks. We used  \textit{librosa}~\cite{mcfee2015librosa} as our audio processing library.
From the resampled audio various handcrafted features are extracted at the frame and segment level, covering frequency-based, structural, statistical and temporal attributes. A segment is the whole instance of one audio recording, while a frame is a chunk (subset) of the whole audio data contained in a segment. %
 A complete list is provided below:
  \\
 
\begin{itemize}

\item \textbf{Duration}: the total duration of the recording after trimming leading and trailing silence. %

\item \textbf{Onset}: the number of pitch onsets (pseudo syllables) is computed from the signals, by identifying peaks from an onset strength envelope,  which is obtained by summing each positive first-order difference across each Mel band~\cite{ellis2007beat}.

\item \textbf{Tempo}: a global acoustic tempo feature is estimated for each recording, which is commonly used in music information retrieval~\cite{ellis2007beat}. It measures the rate of beats which occur at regular temporal intervals. In our context, it is used for its peak detection capabilities.

\item \textbf{Period}: the main frequency of the envelope of the signal.  We calculate the FFT on the envelope and identify the frequency with the highest amplitude from the 4th mode upwards (as the envelope has  non-zero mean).  %

\item \textbf{RMS Energy}: the root-mean-square of the magnitude of a short-time Fourier transform which provides the power of the signal. %

\item \textbf{Spectral Centroid}: the mean (centroid) extracted per frame of the magnitude spectrogram.

\item \textbf{Roll-off Frequency}: the center frequency for a spectrogram bin so that at least 85\% of the energy of the spectrum in this frame is contained in this bin and the bins below.

\item \textbf{Zero-crossing}: the rate of sign-changes of the signal.%

\item \textbf{MFCC}: Mel-Frequency Cepstral Coefficients obtained from the short-term power spectrum, based on a linear cosine transform of the log power spectrum on a nonlinear Mel scale. MFCCs are amongst the most common features in audio processing~\cite{davis1980comparison}. We use the first 13 components.

\item \textbf{$\Delta$-MFCC}: the temporal differential (delta) of the MFCC \cite{librosa}. %

\item \textbf{$\Delta^2$-MFCC}: the differential of the delta of the MFCC (acceleration coefficients)~\cite{librosa}.%
\end{itemize}

For the  features that generate time series (RMS Energy, Spectral Centroid, Roll-off Frequency, and all variants of MFCCs), we extract several statistical features in order to capture the distributions beyond the mean. A complete list is: \textit{mean, median, root-mean-square}, \textit{maximum, minimum, 1st and 3rd quartile, interquartile range, standard deviation, skewness}, and \textit{kurtosis}. In total, there are 477 handcrafted features including the first four segment-level features, four frame-level features represented by their statistics, and three variants of MFCCs with each component represented by its statistics ($4+4\times11 + 3\times13\times11=477$).

\textbf{Features from Transfer Learning.} %
In addition to handcrafted features, we employ \textit{VGGish} to extract audio features automatically~\cite{hershey2017cnn}. \textit{VGGish} is a convolutional neural network that was proposed for audio classification based on raw audio input; the VGGish model was trained using a large-scale YouTube dataset and the learned model parameters were released publicly. We employ it as a feature extractor to transform the raw audio waveforms into embeddings (features), which are then passed to train a shallow classifier. The VGGish pre-trained model first divides data samples into 0.96-sec non-overlapping sub-samples, and for each 0.96 second, it returns a 128-dimensional feature vector. The sampling rate is 16 KHz. We take the mean and standard deviation across the entire segment as the final features, with dimension 256 ($128\times2$). Since VGGish is based only on a spectrogram input, some important characteristics from the temporal domain might get missed in the feature space, which motivates the additional use of a  combination of VGGish with handcrafted features.  Section~\ref{eval} shows that this combination helps in achieving better AUC compared to using solely VGGish or handcrafted features.  %

We obtain a 477-dimensional handcrafted feature vector, a 256-dimensional VGGish-based feature vector, and several combined feature vectors for each modality (cough, breathing) which are up to 733 in dimensions, in total. Each combined feature vector is the concatenation of a subset from the handcraft feature sets and the VGGish-based features.  %
These feature vectors are further reduced by Principal Components Analysis (PCA) retaining a portion of the initial explained variance. More details about the pre-processing are provided in Section~\ref{eval}.

%% file: evaluation.tex
\section{Evaluation}
\label{eval}
We now detail our evaluation of the classification of audio samples as COVID-19 or healthy using features described in Section~\ref{method}. Given the large class imbalance, a subsample of the initially collected dataset (described in Section~\ref{subset-data}) was used. We first describe how data from different modalities were merged and how the dataset was partitioned for the experiments.
To foster reproducibility the experiments\footnote{Experiments source at \url{https://github.com/cam-mobsys/covid19-sounds-kdd20}} 
as well as the Android\footnote{Android app source at: \url{https://github.com/cam-mobsys/covid19-sounds-android-app}} and iOS\footnote{iOS app source located at: \url{https://github.com/cam-mobsys/covid19-sounds-ios-app}} apps code-bases are publicly available. 
Findings and results are discussed in the later part of the section.
\subsection{Experimental setup}
\label{setup}
\textbf{Classification tasks.} Based on the data collection (Section~\ref{data}) we focus on three clinically meaningful binary classification tasks:
\begin{itemize}
    \item \textbf{Task 1}: Distinguish users who have declared they tested \textit{positive for COVID-19} ({\em COVID-positive}), from users who have \textit{not declared a positive test} for COVID-19, have a \textit{clean medical history}, have \textit{never smoked}, have \textit{no symptoms} and, as described in Section~\ref{data}, were in \textit{countries} where COVID-19 was not prevalent at the time ({\em non-COVID}). While we cannot guarantee they were not infected, the likelihood is very small.
    \item  \textbf{Task 2}: Distinguish users who have declared they tested \textit{positive for COVID-19} and have declared a \textit{cough} as a symptom (a frequent symptom in those with COVID, as reported in Figure~\ref{symptoms}), (COVID-positive with cough) from users who have \textit{declared not to have tested positive} for COVID-19, have a \textit{clean medical history}, \textit{never smoked}, were in \textit{countries} where at the time COVID-19 was not prevalent and have a cough as a symptom (non-COVID with cough). 
    \item \textbf{Task 3}:  Distinguish users who have declared they \textit{tested positive for COVID-19} and have declared a \textit{cough} as a symptom (COVID-positive with cough), from users who have \textit{not declared to have tested positive} for COVID-19, are from \textit{countries} where at the time COVID-19 was not prevalent, have reported \textit{asthma} in their medical history and have a \textit{cough} as a symptom (non-COVID with cough). 
\end{itemize}

\begin{figure*}[ht]
    \centering
    \includegraphics
        [trim={0 0.6cm 0 0}, width=1\textwidth]
        {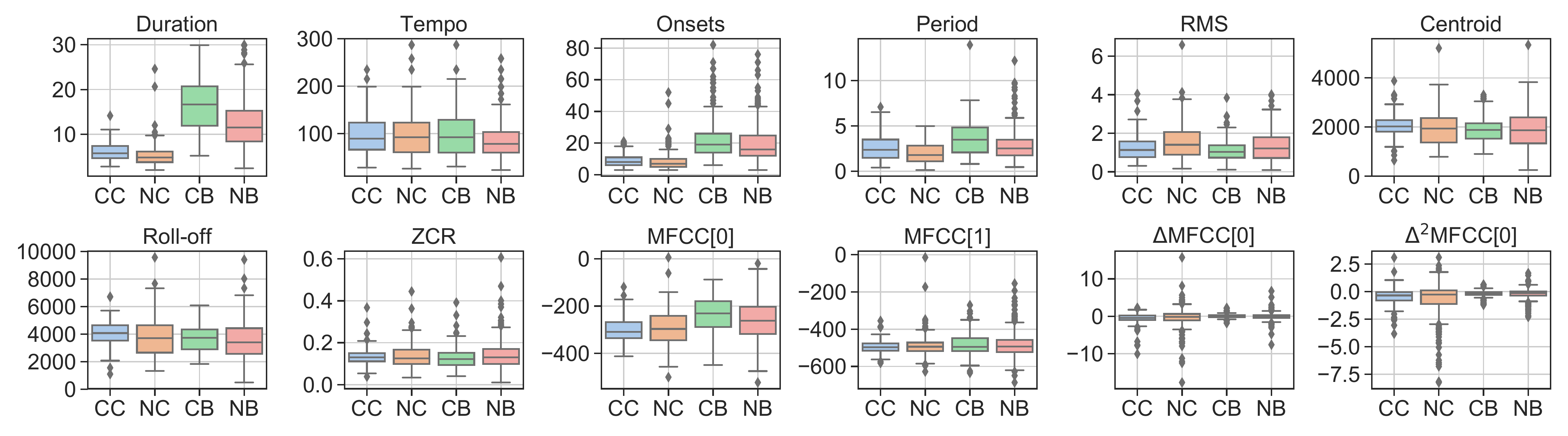}
    \caption{Box plots of the \textit{mean} features of cough and breathing. CC: COVID Cough, NC: Non-COVID Cough, CB: COVID Breathing, NB: Non-COVID Breathing. 
    }
    \label{fig:features_task1}
\end{figure*}

\textbf{Data exploration.} %
As a first step after feature extraction, we examine the differences between the distributions of the  features obtained from cough and breathing broken down by respective class. Given the high dimensionality of the features, we cannot present all distributions, therefore we focus only on the \textit{mean} statistical feature of each feature family (e.g., Centroid is Centroid \textit{mean} here).   The box plots in Figure~\ref{fig:features_task1} show that coughs and breaths from COVID-positive users are longer in total duration, have more onsets, higher periods, and lower RMS, while their MFCC features [1st component and deltas] have fewer outliers.  Across both tasks, the samples from COVID-positive users concentrate more towards the mean of the distributions, whereas the general (healthy) population shows greater span (interquartile range), with the hypothesis being that a (possibly forced) healthy cough and breathing are very diverse. This may also suggest that coughs and breaths are useful sounds for classifying users as COVID or non-COVID.  %

\textbf{Feature ablation studies}. In order to identify which audio
modality (cough or breathing) contributes more to the classification performance, we repeat our experiments with three different audio inputs: %
\textit{only cough}, \textit{only breathing}, and \textit{combined}. To account for the increasing dimensionality of the combined representation and to make a fair comparison, we perform experiments to find the best cut-off value for PCA (see results in next section). The values of explained variance range between [70\%, 80\%, 90\% and 95\%]. In practice, this means that with lower explained variance the classifiers will use fewer input dimensions and vice versa. 
Intuitively, a combined representation might need a more compressed representation than a representation using only coughs or breaths, to prevent overfitting.

\begin{table*}[th]
\begin{tabular}{@{}p{4cm}ccclllll@{}}
\toprule
Task     &Modality          & Samples (users)*& Feature Type        & \multicolumn{3}{c}{Mean $\pm$ std} 
\\ \cmidrule(l){1-7} 
\multicolumn{1}{c}{}   & &   & &ROC-AUC         & Precision       & Recall \\ \cmidrule(l){5-7} 
\multirow{2}{4cm}{1. COVID-positive / non-COVID} &\multirow{2}{*}{Cough+Breath} &\multirow{2}{*}{141 (62) / 298 (220)}&
1 & 0.71 (0.08)  & 0.69 (0.09) & 0.66 (0.14) \\
  &  & &2 & 0.78 (0.07)  & 0.72(0.08) & 0.67(0.11) \\
   &  & &3(A) & \textbf{0.80(0.07)}  & 0.72(0.06) & 0.69(0.11) \\ \hline \\
\multirow{2}{4cm}{2. COVID-positive with \textbf{cough} / non-COVID with \textbf{cough}} &\multirow{2}{*}{Cough}& \multirow{2}{*}{54 (23) / 32 (29)}
&1&  0.65(0.22)  & 0.62(0.20) &  0.69(0.14)  \\
  &  & &2 & 0.82(0.16)  & 0.79(0.16) & 0.71 (0.23) \\
   &  & &3(A) & \textbf{0.82(0.18)}  & 0.80(0.16) & 0.72(0.23) \\ \hline \\

\multirow{2}{4cm}{3. COVID-positive with \textbf{cough} / non-COVID \textbf{asthma cough}} &\multirow{2}{*}{Breath}&\multirow{2}{*}{54 (23) / 20 (18)} 
& 1&  0.76(0.30) &  0.64(0.29) & 0.72(0.31) \\ 
  &&&2  & 0.72(0.16) & 0.77(0.22) & 0.47(0.15)\\ 
 & &&3(B) & \textbf{0.80(0.14)} & 0.69(0.20) & 0.69(0.26) \\ 

\bottomrule
\end{tabular}
\caption{Classification results for the three tasks. *The number of samples before splitting to train/test and downsampling. Logistic Regression results are reported for the first task, while SVMs for the latter two tasks. We report the best modality and representation size for PCA (detailed results for every cutoff are provided in Figure \ref{fig:lineplot}) for each task. Feature Type 1 = Handcrafted with PCA = 0.8 for three tasks, Type 2  = VGGish with PCA = 0.95 for Task 1 and 3, 0.9 for Task2, Type 3 = Handcrafted + VGGish with PCA = 0.95 for Task 1, 0.9 for Task 2, and 0.7 for Task 3. For Type 3, (A) denotes that we use VGGish-based feature plus duration, tempo, onset, and period, (B) for all features except $\Delta$-MFCCs and $\Delta^2$-MFCCs, and (C) for all features. }
\label{cough-table}
\end{table*}

\textbf{User-based cross-validation}. %
We create training and test sets from disjoint user splits, making sure that samples from the same user do not appear in both splits. Note that this does not result in perfectly balanced class splits; however, we downsampled the majority (non-COVID) class when needed. The test set is kept balanced.%

Even then, it is not easy to guarantee that a split selects a representative test-set, so we performed a 10-fold-\textit{like} cross validation using 10 different
random seeds to pick disjoint users in the outer loop (80\%/20\% split), and a hyper-parameter search as inner loop to find the optimal parameters (using the 80\% train-set in a 5-fold cross validation). %
Essentially, this setup resembles a \textit{nested cross-validation} \cite{cawley2010over}.  We conduct extensive
experiments by testing 5400 models (3 tasks × 3 modalities × 10 user splits  × 4 dimensionality reduction cut-offs × 3 feature types  ×  5 hyper-parameter cross-validation runs). We selected several standard evaluation metrics such as the Receiver Operating Characteristic - Area Under Curve (ROC-AUC), Precision, and Recall. We report the average performance of the outer folds (10 user-splits) and the standard deviation. In the following section we report the performance of our three tasks.

\textbf{Sensitivity to demographics.} Including age and sex as one-hot-encoded features in our models (e.g. age group: 40-49 years old) did not improve or worsen the results substantially ($<$ $\pm$ 2 AUC).

\subsection{Distinguishing COVID-19 users from healthy users}
Table~\ref{cough-table} reports the results of the classification for the three tasks described above. For each task, we report the best  results, which might have been obtained using either a single modality (cough or breathing sounds) or a combination of both modalities.     The first row reports classification results for Task 1: the binary classification task of discriminating users who declare having tested positive for COVID-19 ({\em COVID-positive}), from users who answered no to that question ({\em non-COVID}). 

The metrics  show that there seem to be some discriminatory signals in the data indicating that user coughs combined with breathing could be a good predictor when screening for COVID-19. In particular, the AUC for this task is at 80\% while precision and recall are around  70\%. Compared to the other tasks (Task 2 and 3), Task 1 has the lowest standard deviations across the user-splits, mostly due to the larger data size. We applied a very simple classifier (Logistic Regression) and that the data is perhaps too limited in size to obviate the noise and diversity introduced by our crowdsourced data gathering (e.g., differences in microphones, surrounding noises, ways of inputting the sounds). Nevertheless these results give us confidence in the power of this signal. We also observed that handcrafted features when combined with features learnt from VGGish provide better results than handcrafted or transfer learning features alone, which shows the promise of using transfer learning in our analysis.

\subsection{Distinguishing COVID-19 coughs from other coughs} 

The second row of Table~\ref{cough-table} describes the binary classification of users who reported testing positive for COVID-19 and also declared a cough in the symptom questionnaire, and a similar number of users who said they did not test positive for COVID-19 but declared a cough (Task 2). The best result shows an AUC of 82\%. Precision for this task is at 80\%, showing that cough sounds are able to distinguish COVID-19 positive users quite well. Recall is slightly lower (72\%), meaning that this model casts a good but rather specialized net: it does not detect every COVID-19 cough, but many of them.  Nevertheless, the size of the data, as well as the relatively high standard deviations compared to Task 1, renders this result preliminary. 

We also compared COVID-19 users with a cough, described above, to users who said they did not test positive for COVID-19 but reported asthma and declared a cough. The last row of Table~\ref{cough-table} shows an AUC of 80\%. While the recall is acceptable, precision for this task is also high, like for the other two tasks.  It is interesting to see that breathing sounds serve as more powerful signals to discriminate users in this task.  
We have further evaluated the utility of data augmentation for Task 2 and 3 to improve performance (Section \ref{augmentation}).

%% file: discussion.tex
\section{Discussion and Conclusions}
We have presented an ongoing effort to crowdsource respiratory sounds and study how such data may aid COVID-19 diagnosis. 
These results only scratch the surface of the potential of this type of data; while our results are encouraging, they are not as solid as would be necessary to constitute a standalone screening tool.
We have, for the moment, limited ourselves to the use of a subset of the data collected, to manage the fact that the proportion of COVID-19 positive users is small. We also have no ground truth regarding health status, and so took users from countries where COVID-19 was not prevalent at the time as likely to be truly healthy when self-reporting as such (however, this limited our dataset further). We are in the process of collecting more data and discussing how this crowdsourced endeavor could be complemented by a controlled one, where we deliberately collect samples only from users who have had a positive or negative COVID test as ground truth. 
This will allow analysis of a larger dataset, possibly with more advanced machine learning (e.g., deep learning). We are extending our study to voice sounds, which we have already collected. Vocal patterns, alongside breathing and cough, could give useful additional features for classification.

While we have shown a limited investigation of the difference between cough sounds in COVID-19 and asthma, our dataset also includes users with other respiratory pathologies, and we hope to study this further to investigate how distinguishable COVID-19 is in this respect.

The mobile app reminds users to provide samples every couple of days: as a consequence we have a number of users for whom we could study the progression of respiratory sounds in the context of the disease. This is very relevant for COVID-19, and something we have not yet investigated in the current work.

Finally, our current app does not offer medical advice and only collects data; while we hope that the models developed from this data will be useful in disease screening, we are aware of the challenges involved in giving medical advice to users and the debates that this generally sparks~\cite{ForbesSingh}.

%% file: appendix.tex
\onecolumn

\section{Supplementary Material}

\subsection{The impact of modalities and dimensionality} 

\begin{figure*} [h]
    \centering
    \begin{subfigure}{.3\textwidth}
        \centering
        \includegraphics[scale=0.26]{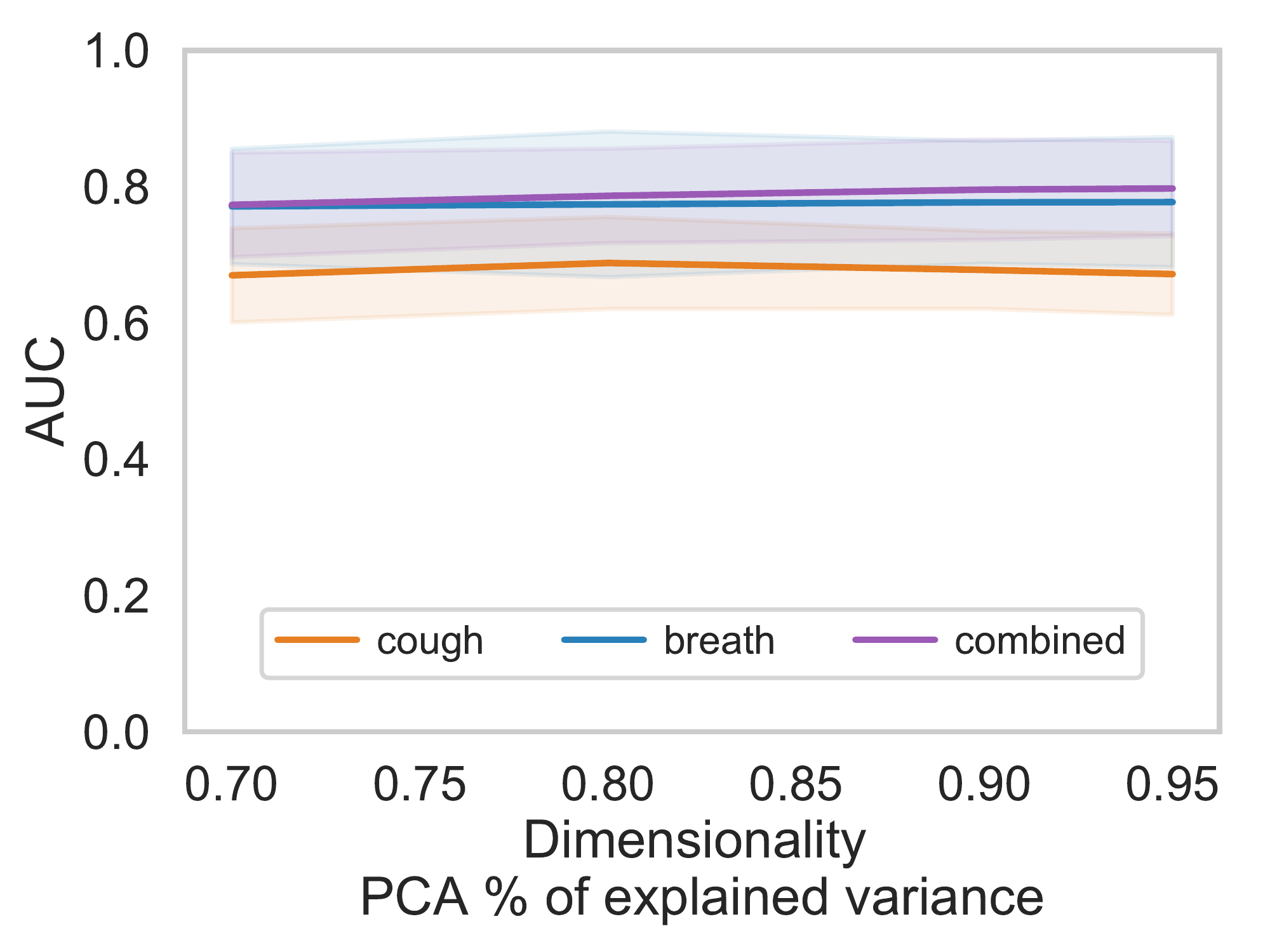}
        \caption{COVID-positive / non-COVID\vspace{12pt}}
        \label{fig:lineplot:a}
    \end{subfigure}
    \hspace{2pt}
    \begin{subfigure}{.3\textwidth}
        \centering
        \includegraphics[scale=0.26, angle=0]{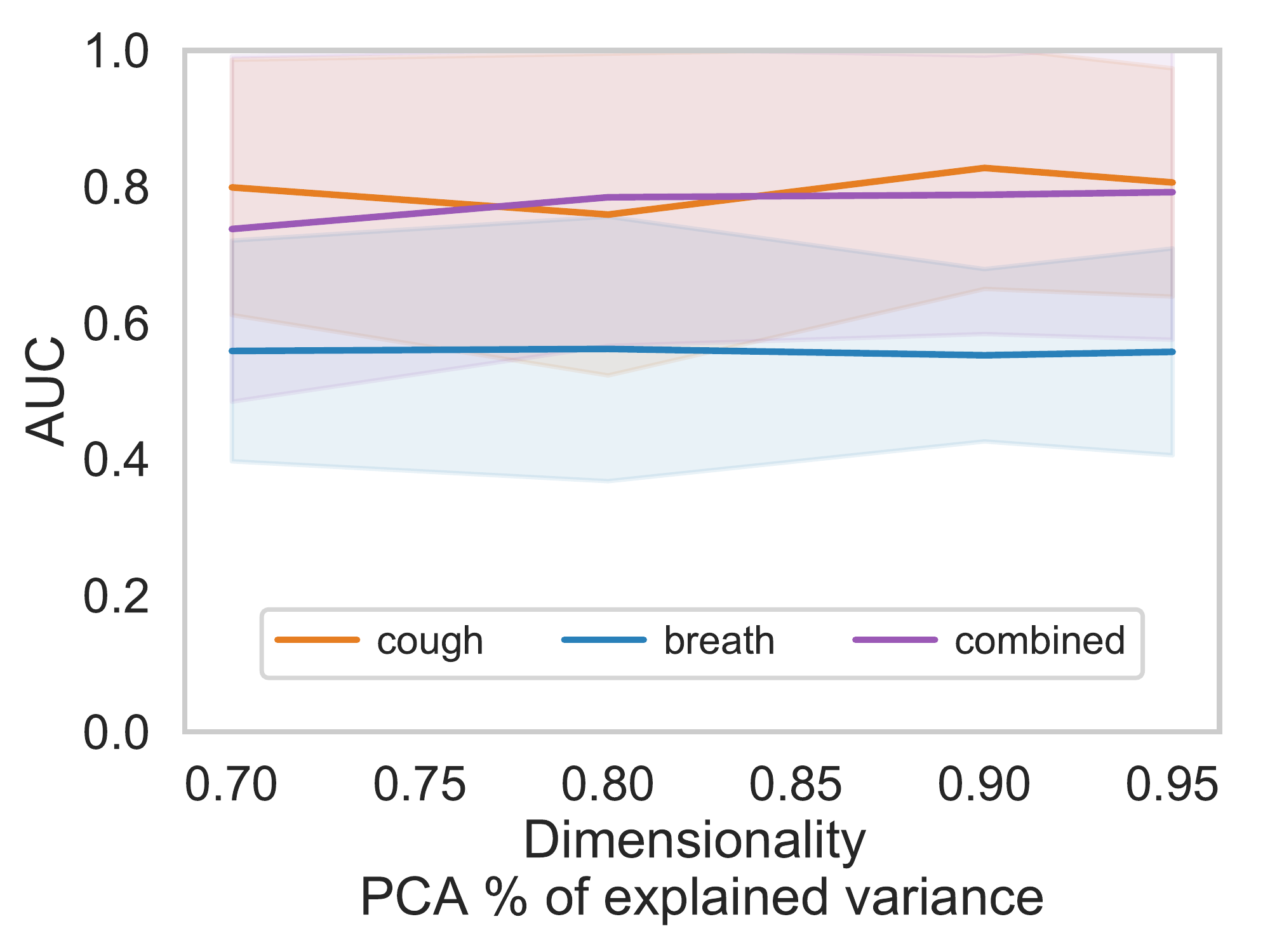}
        \caption{COVID-positive with \textbf{cough} / non-COVID with \textbf{cough}}
        \label{fig:lineplot:b}
    \end{subfigure}
    \hspace{2pt}
    \begin{subfigure}{.3\textwidth}
        \centering
        \includegraphics[scale=0.26, angle=0]{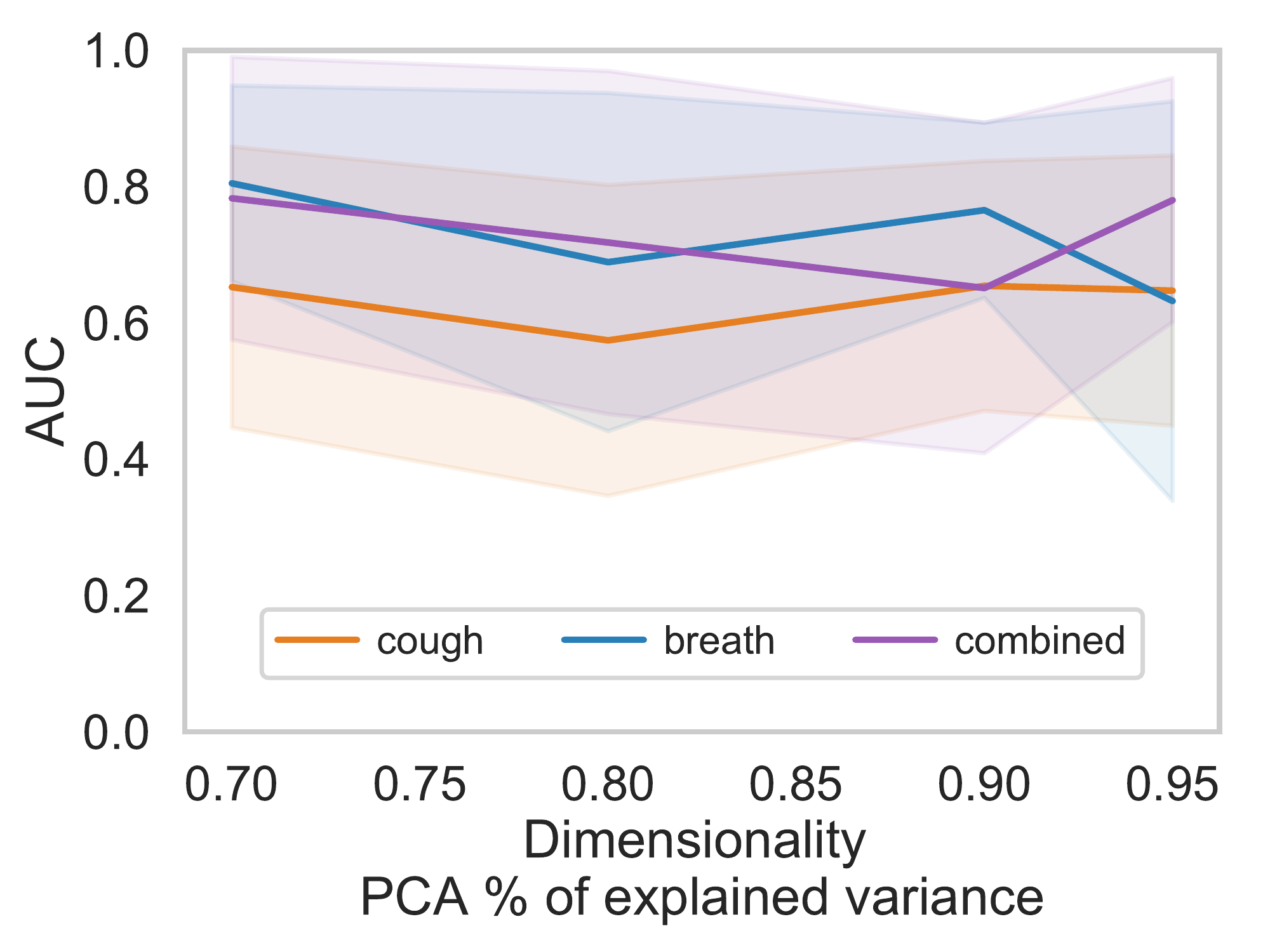}
        \caption{COVID-positive with \textbf{cough} / non-COVID with \textbf{cough}}
        \label{fig:lineplot:c}
    \end{subfigure}
      \vspace{-3mm}
    \caption{The effect of combining different sound modalities (cough, breathing) and the size of the feature vector dimension on overall performance (AUC $\pm$ std in shaded areas). We note that Tasks 2 and 3 (b,c) overfit more due to sample sizes.
    } 
    \label{fig:lineplot}
  \vspace{-3mm}
\end{figure*}

Our overall results suggest that different modalities are useful for different tasks. Here we analyze the role of each individual modality and the combination in more detail for the three tasks.   Figure~\ref{fig:lineplot:a} shows that for Task 1, cough alone (at least for the simple features used) performs reasonably well (AUC around 70\%); however in combination with breathing sounds, it achieves the highest AUC and lowest standard deviation for the task.  The dimensionality size is not highly significant here, however, the combination seems to improve with more features. %
For Task 2, in Figure~\ref{fig:lineplot:b} we observe a different trend where cough is more precise than   breathing or their combination. This finding manifests from the nature of the task itself as data pertains to the people who actually had a cough. This also highlights the fact that cough can be an important biomarker to differentiate between COVID and non-COVID users. Although the combined feature set achieves better AUC with lower dimensionality (PCA 0.8), the cough modality seems to improve by using more features. This is expected, due to the different feature sizes. %
Lastly, Task 3 (Figure~\ref{fig:lineplot:c})  follows similar trends, with overfitting becoming more apparent in higher dimensions due to smaller sample size. Here, the breathing modality outperforms the other modalities, and its performance  is highest at PCA 0.7  suggesting that breathing is an important biomarker to distinguish COVID sounds from asthma.  %
\vspace{-4mm}

  \vspace{-3mm}
\subsection{Data augmentation} 
\label{augmentation}
To counter the small amount of control data available for Tasks 2 and 3, we augmented the negative class (non-COVID) for these two tasks using three standard audio augmentation methods~\cite{schluter2015exploring}: amplifying the original signal (1.15 to 2 times, picked using a random number), adding white noise (without excessively impacting signal to noise ratio), and changing pitch and speed (0.8 to 0.99 times). We made sure not to distort the original signal significantly: we manually inspected and listened to the audio before and after performing data augmentation. We applied each method twice to the original samples to obtain six times as many. Specifically, we increased the number of samples for `non-COVID with cough' and `non-COVID asthma with cough'. Note that we used augmented samples only for training (the test set was kept intact). The results are shown in Table~\ref{cough-augmented-table}. We observe that the performance for all the metrics improved. There is almost a 10\% increase in both the AUC (nearly 90\%) and recall, with a slightly smaller standard deviation, compared to results in Table~\ref{cough-table}. A very high AUC shows that audio sounds have high discriminatory power to distinguish COVID vs non-COVID and asthma patients. With the much improved recall our model is also able to recognize almost  all the {\em COVID coughs}. This is clinically important, since our aim is to identify COVID-19 positive cases; misclassifying some healthy users is acceptable as these can be identified in a second stage of testing. %

\vspace{-2mm}
\begin{table*}[h]
\begin{tabular}{@{}p{4cm}ccclllll@{}}
\toprule
Task     &Modality          & Samples (users)*& Feature Type        & \multicolumn{3}{c}{Mean $\pm$ std} 
\\ \cmidrule(l){1-7} 
\multicolumn{1}{c}{}   & &   & &ROC-AUC         & Precision       & Recall \\ \cmidrule(l){5-7} 
\multirow{2}{4cm}{2. COVID-positive with \textbf{cough} / non-COVID with \textbf{cough}} &\multirow{2}{*}{Cough}& \multirow{2}{*}{54 (23) / 32 $\times$ 6 (29)} 
& 1& 0.58(0.22) & 0.52(0.19)& 0.85(0.30) \\ 
  &  & &2 & 0.73(0.20)  & 0.57(0.16) & 0.92(0.24) \\
   &  & &3(C) & \textbf{0.87(0.14)}  & 0.70 (0.15) & 0.90 (0.18) \\ \hline \\

\multirow{2}{4cm}{3. COVID-positive with \textbf{cough} / non-COVID \textbf{asthma with cough}} &\multirow{2}{*}{Breath}&\multirow{2}{*}{54 (23) / 20 $\times$ 6 (18)} 
& 1& 0.77(0.18) & 0.68(0.09)& 0.90(0.16) \\ 
  &&&2  & 0.88(0.15) & 0.63(0.22)& 0.82(0.32) \\ 
 & &&3(B) & \textbf{0.88(0.12)} & 0.61(0.22)& 0.81(0.31) \\ 

\bottomrule
\end{tabular}
\caption{Classification results with data augmentation for Tasks 2 and 3. Task 2: PCA = 0.8 for Type 1 and 0.7 for Type 2 and 3; Task 3: PCA = 0.95 for Type 1 and 0.9 for Type 2 and 3.}
\label{cough-augmented-table}
\end{table*}